\documentclass[letterpaper, 10 pt,conference]{ieeeconf}
\IEEEoverridecommandlockouts
\usepackage[makeroom]{cancel}
\usepackage{amsmath,amssymb,amsfonts}
\usepackage{abraces}
\usepackage{algorithmic}
\usepackage{graphicx}
\usepackage{textcomp}
\let\labelindent\relax
\usepackage{enumitem}

\usepackage{xcolor}
\def\BibTeX{{\rm B\kern-.05em{\sc i\kern-.025em b}\kern-.08em
    T\kern-.1667em\lower.7ex\hbox{E}\kern-.125emX}}

\usepackage{stmaryrd} 

\usepackage{amsthm}

\newtheorem{theorem}{Theorem}
\newtheorem{lemma}{Lemma}
\newtheorem{corollary}{Corollary}

\newtheorem{remark}{Remark}
\newtheorem{assumption}{Assumption}

\usepackage{mathtools,bm}

\usepackage{tikz}
\usetikzlibrary{positioning, arrows.meta,calc}
\usepackage{pgfplots}

\usepackage{subfigure}

\usepackage[demo]{adjustbox}
\usepackage{float, caption, makecell, cellspace}
\usepackage{newtxtext, newtxmath} 
\usetikzlibrary{decorations.pathreplacing,angles,quotes}

\begin{document}
\setlist[itemize]{leftmargin=*}

\title{Excitability of High-Frequency Positive Feedback}

\author{Gustave Bainier$^{1}$, Arthur Fyon$^{1}$, Christian Fernandez Lorden$^{1}$, Alessio Franci$^{1,2}$
\thanks{$^{1}$ Dept. of Electrical Engineering and Computer Science, University of Liège, 4000 Liège, Belgium; {\tt\small gustave.bainier@uliege.be} $^{2}$ WEL Research Institute, 1300 Wavre, Belgium, 
       }%
\thanks{This work was supported by the Belgian Government through the Federal Public Service Policy and Support.}
}

\maketitle

\begin{abstract} What is the threshold of an excitable system, i.e. the criterion that distinguishes rest from excitation? Previous research has shown that this question is far from trivial to answer. This paper studies the threshold of a simple excitable primitive in which excitability arises from the positive feedback interconnection of a first-order high-pass filter and a static, strictly monotone and strictly convex activation function. This threshold is a simple inequality on the difference between the instantaneous input and the motif's internal memory. While this result may help neuro-engineers characterize spiking in biological and artificial neurons, we argue that excitability is not always a desirable property, and that our threshold may also be relevant to safety considerations in fragile systems with rate-induced tipping points. We illustrate our framework with three examples: the FitzHugh-Nagumo model of neuronal dynamics, a model of stop-loss cascades in the stock market, and a social contagion model.
\end{abstract}

\section{Introduction}

\subsection{The robust fragility of excitable systems}

\begin{figure}
    \centering
    \includegraphics[width=0.95\linewidth]{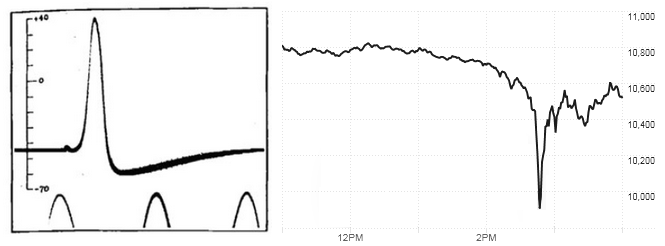}
    \caption{\small \textbf{Left}: First intracellular recording of an action potential captured in the squid giant axon~\cite{Hodgkin1952}. \textbf{Right}: Flash crash of the Dow Jones on May 6th, 2010~\cite{sec_cftc_2010_flash_crash}.}
    \label{fig:schizo}
\end{figure}

Excitability is a property by which a system responds in an all-or-none fashion to the transient of an input signal. In neurophysiology, it is the well-studied spiking mechanism of neurons, where a large, stereotyped action potential is produced in response to a sufficiently large and/or sharp input current~\cite{Sepulchre2018} (Fig.~\ref{fig:schizo}, Left; Fig.~\ref{fig:noEnergetic}).

\begin{figure}
    \centering
    \includegraphics[width=0.95\linewidth]{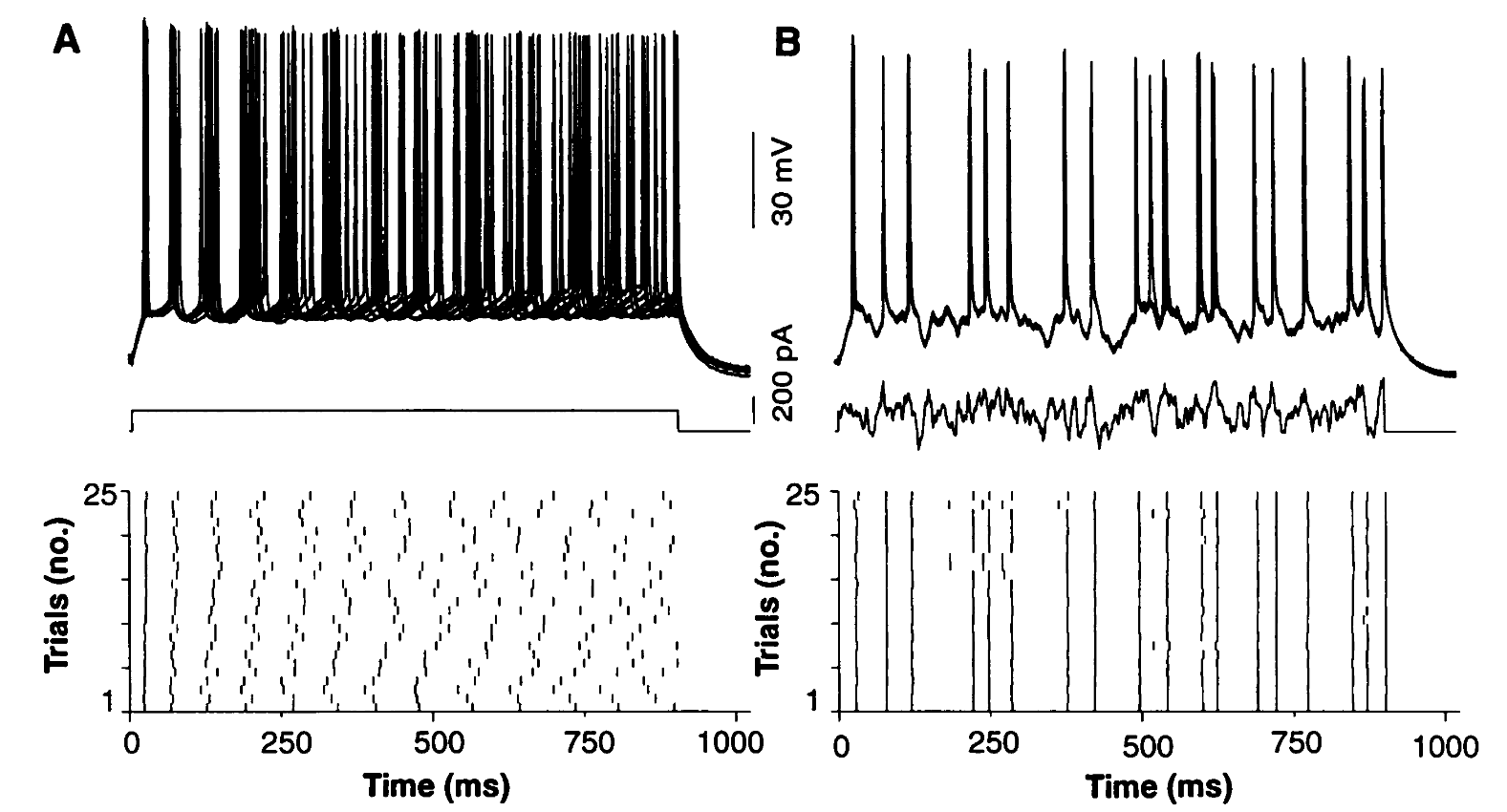}
    \caption{\small Reproduced from~\cite{Mainen1995}. \textbf{Left}: A constant input yields unreliable spike timing across trials. \textbf{Right}: A fluctuating input yields reliable spike timing across trials.}
    \label{fig:reliable}
\end{figure}

From both the neurophysiological and neuromorphic engineering standpoint, excitability is a desirable property. Excitability makes the spike timing ultra-sensitive to the transient of the input signal rather than to the neuron's parametric uncertainties, so the production of action potentials is reliable under neuronal heterogeneity and/or neuromorphic hardware mismatch~\cite{Mainen1995,Sepulchre2026,Loris2026} (Fig.~\ref{fig:reliable}). As such, excitability enables the robust encoding of information in the spike timing, rather than merely in the spiking frequency~\cite{Denve2016}. This mechanism bridges the continuous, analog world with the discrete, digital, event-based one: smooth, continuous-time analog signals are robustly reshaped into a discrete set of sparse timed events~\cite{Sepulchre2022}. 

Although excitability is desirable in neural systems, that may be the exception rather than the rule. Indeed, the excitability property is also a defining feature of \textit{fragile systems}, that may lead to catastrophic failures. A blackout propagates across a continent in minutes; a systemic financial collapse breaks the world economy through cascading bank failures; the weakening of the Atlantic Meridional Overturning Circulation (AMOC) is self-amplified through global warming. All of these examples involve a fast self-reinforcing loop whose activation depends on the rate of change of a signal~\cite{Ritchie2023,Brunnermeier2008}. The same property that makes neurons spike timing reliable could be making other systems \textit{fail} reliably. In the tipping-point literature, a system's ultra-sensitivity to a rate of change is also known as rate-induced tipping~\cite{Ashwin2012}. Excitable systems are a natural class in which rate-induced tipping occurs, and in this paper, we call such systems \textit{excitably fragile}. 

A surprisingly large range of systems may qualify as excitably fragile. In the Earth system, we find evidence of rate-induced tipping in the carbon emissions from permafrost soils~\cite{OSullivan2023}. In mechanical systems, stick-slip friction, which is observed in machine-tool chatter or in earthquakes, produces slip events with a spike-like signature~\cite{Berman1996}. In power systems, cascading blackouts have been attributed to rate-induced tipping~\cite{Nedic2006,Ritchie2023}. Related phenomena may arise in electronics, where CMOS latch-up and thermal runaway in power devices are already plausible candidates~\cite{Troutman1983,Spirito2002}. Financial markets provide other speculative examples, with flash crashes and liquidity spirals exhibiting similar behaviors~\cite{Brunnermeier2008,kirilenko2017flash,Fosset2020} (Fig.~\ref{fig:schizo}, Right).

Whether viewed from a neurophysiological, neuroengineering, tipping-point, or bifurcation-control perspective, the capacity of an excitable systems to blow up in response to a peculiar transient would benefit from a well-defined notion of threshold.~\cite{Ritchie2023,Ashwin2012,Mitry2013,Sepulchre2025,Bifurcation2003} (Fig.~\ref{fig:noEnergetic}). This work addresses this question for a newly introduced motif, where excitability is understood as a high-frequency positive feedback loop.

\subsection{A dynamical threshold for high-pass excitability}


An elementary motif of neural excitability has recently been proposed in neuroengineering~\cite{Bainier2026_poster}. Succinctly, it consists of the positive feedback interconnection of a high-pass filter and a high-gain sigmoid (Fig.~\ref{fig:blocs}, Top). The motif can reproduce both Hodgkin's Type~I and Type~II excitability~\cite{Izhikevich2006}, depending on the shape of the sigmoid. Furthermore, it offers a frequency-based point of view on the mechanism: the motif produces a spike through a positive feedback amplification of the high-frequencies components of its input signal.

The analysis carried out in this paper is strictly focused on obtaining an exact spike-initiation threshold. For this reason, this work further simplifies this recently introduced motif. We consider a strictly monotone and convex activation function rather than a sigmoid, so the local spike-producing bifurcation remains, but no inflection point produces a limit cycle terminating the spike. This simplification is akin to that used in hybrid neuron models, where the spike termination is replaced by an artificial reset rule. Examples include the quadratic/exponential integrate-and-fire models and their adaptive extensions, or the Izhikevich model~\cite{Izhikevich2006}. Here, however, the simplification serves a theoretical rather than a computational purpose. It allows us to focus the analysis on the excitability threshold, which we can then define as the point at which the inverse motif ceases to be well-defined (Fig.~\ref{fig:blocs}, Bottom).

\begin{figure}
    \centering
    \includegraphics[width=\linewidth]{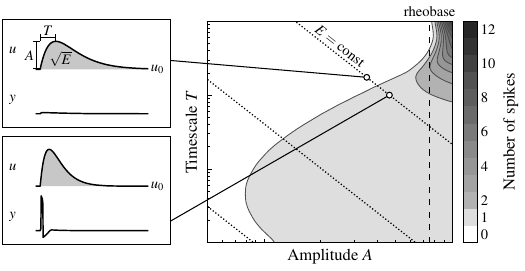}
    \caption{\small Inspired by~\cite{Sepulchre2018}. Simulation of a FitzHugh-Nagumo neuron model. The spiking threshold is neither static, neither easily captured by simple energetic arguments.}
    \label{fig:noEnergetic}
\end{figure}

Characterizing the threshold of an excitable system is generally considered to be a difficult task~\cite{Izhikevich2006}. In neurophysiology, spike generation is associated with a bifurcation in the state space of the neuron model at a static input level: the rheobase. Although the rheobase is the static threshold of spike generation, it is well documented that spikes can be triggered by inputs that remain at all times below this static threshold, and a \textit{dynamical threshold} must be characterized in order to anticipate the production of a spike ~\cite{Izhikevich2006,Mitry2013} (Fig.~\ref{fig:noEnergetic}). While this dynamical threshold is not easily captured by simple energetic arguments, a recent work in neuromorphic control successfully introduced an history-dependent energy-based threshold, by computing the minimal external supply that is required to trigger a spike~\cite{Sepulchre2025}. This work is more general than ours, and it requires solving an optimal control problem. By contrast, thanks the simplicity of the motif considered here, our threshold is much simpler. It consists of a simple inequality on the difference between the instantaneous input and the motif's internal memory. This threshold naturally accounts for both excitatory and post-inhibitory spikes.

Beyond neuroscience and neuroengineering, if the motif of excitably fragile systems is akin to the one introduced in this paper, then this bound may constitute a generic tipping-point result. Because the blocks involved are extremely elementary, we argue that our proposed motif is very likely to be found in rate-induced tipping point systems. Positive feedback high-pass filtering emerges naturally in systems that are excited by their own rate of change. We describe two examples of such excitably fragile systems at the end of the paper: stop-loss cascades in stock market dynamics, which have notably been implicated in amplifying the 2010 flash crash~\cite{sec_cftc_2010_flash_crash} (Fig.~\ref{fig:schizo}, Right), and social contagion, when a behaviour spreads in response to a recent increase in participation.

The contributions of this paper are the following.
\begin{enumerate}
    \item In Section~\ref{sec:motif}, we introduce the minimal high-frequency positive feedback excitability motif, as well as its stable inverse.
    \item In Section~\ref{sec:threshold}, we derive the explicit dynamical threshold above which the motif blows up, and below which the motif remains stable.
    \item In Section~\ref{sec:examples}, we provide three examples fitting the proposed excitable motif: the FitzHugh-Nagumo model of a neuron, a stop-loss cascade model in the stock market, and a social contagion model. 
\end{enumerate}

Finally, Section~\ref{sec:conclusion} concludes the paper and offers some perspectives.

\section{The excitable motif and its stable inverse}\label{sec:motif}

\subsection{The high-pass excitable motif}

\begin{figure}[t]
    \centering
    \includegraphics[width=0.85\linewidth]{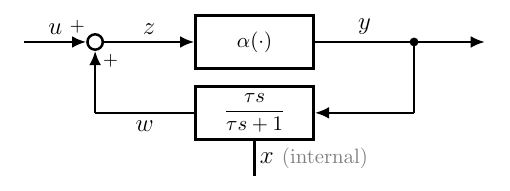} 
    \includegraphics[width=0.85\linewidth]{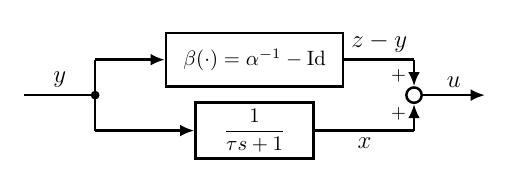}
    \caption{\small \textbf{Top}: Block diagram of the excitable motif... \textbf{Bottom}: ...and its stable inverse.}
    \label{fig:blocs}
\end{figure}

The excitable motif studied in this paper can be described as the positive feedback interconnection between a static activation function $\alpha$ and a high pass filter $H(s)$ (Fig.~\ref{fig:blocs}, Top). For slow-varying inputs, the system is predominantly feedforward, as the high-pass filter in the feedback loop suppresses constant signals. In contrast, for fast-varying inputs, the positive feedback loop becomes active and may lead to a fast blow-up.

\begin{assumption}[Activation function]\label{ass:activation} The activation function $\alpha\in \mathcal{C}^2(\mathbb{R}, \mathbb{R}_{> 0})$ is taken strictly increasing and strictly convex. Moreover, it is assumed that:
\begin{subequations}
\begin{align}
    \lim_{x \to -\infty} \alpha(x)=0,&\qquad \lim_{x \to +\infty} \alpha(x)=+\infty,&\\
    \lim_{x \to -\infty} \alpha'(x)=0,&\qquad \lim_{x \to +\infty}  \alpha'(x)>1.&
\end{align}
\end{subequations}
\end{assumption}

A direct consequence of Assumption~\ref{ass:activation} is that the inverse maps $\alpha^{-1}\in \mathcal{C}^{1}(\mathbb{R}_{>0},\mathbb{R})$ and $(\alpha')^{-1} \in \mathcal{C}^0((0,1],\mathbb{R})$ are well-defined and increasing.

\begin{assumption}[High-pass filter] \label{ass:highpass} The high-pass filter is first order:
     \begin{equation}
         H(s)\triangleq \frac{\tau s}{\tau s +1}=1-\frac{1}{\tau s +1}.
     \end{equation}
     Its internal state is denoted $x$, with $x(0)=\alpha(u(0))$.
\end{assumption}

We denote $u$ the input signal, $y$ the output signal, and $w$ the feedback from the high-pass filter. The full excitable motif reads:
\begin{equation} \label{eq:motif}
    y=\alpha(u+w), \quad w=H(s)y,
\end{equation}
with state-space representation:
\begin{subequations}
    \begin{align}
        y&=\alpha(u+y-x), \label{eq:sys1}\\
        \tau\dot{x}&=-x+y, \label{eq:sys2}
    \end{align}
\end{subequations}
where $x \triangleq y-w = \frac{1}{\tau s +1}y$ is the internal state of the high-pass filter. This system is implicitly defined and possibly multi-valued, hence it is acausal and admits no admissible physical realization in general. However, it may be seen as the singular limit of a causal system, where \eqref{eq:sys1} is replaced by
\begin{equation} \label{eq:regul}
    \varepsilon \dot{y}=-y+\alpha(u+y-x),
\end{equation}
with $y(0)=x(0)=\alpha(u(0))$ and $\varepsilon \to 0^+$. Whenever $\alpha$ is a sigmoid, this is a rewriting of the FitzHugh-Nagumo neuron model (see Section~\ref{sec:ex_FHN}). If we introduce the variable $z\triangleq u+w = \alpha^{-1}(y)$, \eqref{eq:motif} can also be re-written as
\begin{equation}
    \tau(1-\alpha'(z))\dot{z}+z=\tau \dot{u}+u,
\end{equation}
where it becomes intuitive that the system remains stable as long as the left-hand side is Hurwitz, i.e. as long as $\alpha'(z)<1$.

\subsection{The inverse motif}

In their seminal paper introducing the electrical dynamics of neurons, Hodgkin and Huxley identified their state-space model through a series of voltage-clamp experiments, effectively computing the inverse model of a squid giant axon membrane~\cite{Hodgkin1952}. Indeed, while the direct model of a neuron (current-to-voltage) is excitable, and is therefore unstable, by contrast, its inverse model (voltage-to-current) is typically feedforward and stable, which greatly facilitates its identification and analysis~\cite{Burghi2019-tf,Burghi2021}. The inverse model of a neuron generally contains both a fast positive branch and a slow negative branch. This observation has motivated the mixed-feedback principle, which suggests that excitability always results from the combination of a fast positive feedback and a slow negative feedback~\cite{Sepulchre2018}.

Inspired by this perspective, we notice that our motif also possesses a stable feedforward inverse  (Fig.~\ref{fig:blocs}, Bottom). Its state-space model reads:
\begin{subequations}
    \begin{align}
        u &= \beta(y)+x, \label{eq:inv_sys1}\\
        \tau \dot{x}&=-x+y, \label{eq:inv_sys2}
    \end{align}
\end{subequations}
where $\beta \triangleq \alpha^{-1}-\mathrm{Id}$.

While our proposed motif of excitability follows the mixed-feedback principle, since the high-pass filter may be understood as a direct (hence, fast) feedthrough minus a slow low-pass filter, we instead decompose our inverse model into a static path and a dynamical one. The dynamical path~\eqref{eq:inv_sys2} accounts for the internal memory of the system, and the static path consists of a concave activation function $\beta$. This $\beta$ bloc is the inverse of the instantaneous positive feedback loop of high-pass motif (Fig.~\ref{fig:collapse}). The unique maximum of $\beta$ is obtained at $\alpha'(z)=1$, as proved hereafter.

\begin{lemma}[Maximum of $\beta$] The map $\beta \in \mathcal{C}^1(\mathbb{R}_{>0},\mathbb{R})$ is strictly concave and attains a maximum at the unique point where $\alpha'(\alpha^{-1}(y))=1$.
    \begin{subequations}
        \begin{align}
        \arg\max \beta &= \alpha\big((\alpha')^{-1}(1)\big),\\ 
        \max \beta &= (\alpha')^{-1}(1)-\alpha\big((\alpha')^{-1}(1)\big).
        \end{align}
    \end{subequations}
\end{lemma}
\begin{proof} Since $\alpha$ is a strictly convex and strictly increasing bijection from $\mathbb{R}$ onto $\mathbb{R}_{>0}$, by the inverse function theorem, $\beta$ is $\mathcal{C}^1$ on $\mathbb{R}_{>0}$ with $\beta'(y)=1/\alpha'(\alpha^{-1}(y))-1$. As $\alpha'$ is strictly increasing and $\alpha^{-1}$ is increasing, $\beta'$ is strictly decreasing, so $\beta$ is strictly concave. Finally, since $\alpha'$ increases continuously from $0$ to a limit greater than $1$, $\beta'$ has a unique zero at $\alpha'(\alpha^{-1}(y))=1$, which maximizes $\beta$.
\end{proof}

This inverse model provides a new lens through which to investigate excitability. In the singular limit under consideration, the system explodes whenever its inverse becomes ill-defined. If the measured values of $u$ and $x$ at a given time are such that no value of $y$ can satisfy \eqref{eq:inv_sys1}, i.e. if $u-x>\max\beta$, then the system blows up. This idea is made rigorous in the next section.


\section{Thresholding the motif}\label{sec:threshold}

\begin{figure}[t]
    \centering
    \includegraphics[width=0.85\linewidth]{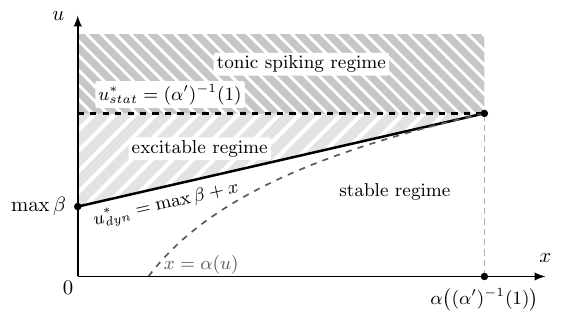}
    \caption{\small Dynamic threshold $u^*_{dyn} = \max\beta + x$ (Theorem~\ref{thm:exact_internal}) and static threshold $u^*_{stat} = (\alpha')^{-1}(1)$ (Corollary~\ref{cor:rheobase}) as functions of $x$. For $u^*_{dyn} < u  \leq u^*_{stat}$, the spike produced is isolated: this is the excitable regime. For $u > u^*_{stat}$, the system is past the bifurcation: this is the tonic spiking regime. The curve $x=\alpha(u)$ represents the equilibria in the stable regime}
    \label{fig:thresholds}
\end{figure}

Building on the intuition gained from the inverse motif, we now demonstrate the exact dynamical threshold of our primitive: $u-x=\max\beta$. Above it, the system blows up; below it, the system remains stable.

\begin{theorem}[Internal state threshold]\label{thm:exact_internal}Assume $u(0)\leq(\alpha')^{-1}(1)$ and let $T\triangleq\inf\{t\geq0:\ u(t)-x(t)>\max\beta\}$. If $T=+\infty$, then the excitable motif never blows up. Otherwise it blows up at time $T$.
\end{theorem}
\begin{proof} Let $f(y)\triangleq-y+\alpha(u+y-x)$ for $y>0$, so $\varepsilon\dot y=f(y)$. Since
$f'(y)=\alpha'(u+y-x)-1$ is strictly increasing and changes sign, we easily find:
\begin{subequations}
\begin{align}
    \arg\min f &=  (\alpha')^{-1}(1)-(u-x), \\
    \min f &=(u-x)-\max\beta .
\end{align}
\end{subequations}
If $T<+\infty$, then at time $T$, we have $\min f>0$, so $\varepsilon\dot y\geq\min f>0$ and $\lim_{\varepsilon\to 0^+}\dot y(T)=+\infty$, so the motif blows up.

If $T=+\infty$, we have $u-x\leq\max\beta$, which is equivalent to $f(\alpha((\alpha')^{-1}(1)))\leq0$. Since $0\leq y(0) \leq \alpha((\alpha')^{-1}(1))$ and $y$ is continuous, $\{y\leq \alpha((\alpha')^{-1}(1))\}$ is positively invariant, so $y$ remains in $[0, \alpha((\alpha')^{-1}(1)]$. Then \eqref{eq:sys2} gives $\dot x\geq 0$ at $x=0$ and $\dot x\leq 0$ at $x=\alpha((\alpha')^{-1}(1))$, so $x$ remains in $[0,\alpha((\alpha')^{-1}(1))]$. Both $x$ and $y$ stay bounded, so the motif never blows up.
\end{proof}

\begin{remark}
    In order to achieve the condition $u-x>\max\beta$, the variable $z\triangleq u+w$ must reach the threshold $\alpha'(z)=1$ from below. Since $z\triangleq u+w$, the threshold $\alpha'(z)=1$ is expressed in terms of the input $u$ and of the feedback $w$, that is, the output of the high-pass filter rather than its internal state $x$.
\end{remark}

At rest, the state variables sit at $x\equiv y \equiv \alpha(u)$ for a constant input $u$. The dynamic threshold then reduces to a static one, which is the rheobase of the motif. In the proposed motif, the rheobase exactly corresponds to the point at which the linearized loop gain becomes greater than one, where the loop becomes unstable.

\begin{corollary}[Rheobase]\label{cor:rheobase}
    Assume the input $u$ is constant, then the
    motif is unstable, if and only if $u>(\alpha')^{-1}(1)$. It will blow up if infinitesimally disturbed.
\end{corollary}
\begin{proof}
    Linearizing \eqref{eq:regul}, \eqref{eq:sys2}, and taking $\varepsilon\to0^+$, we easily find that the equilibrium $x\equiv y \equiv \alpha(u)$ is unstable if and only if $\alpha'(u)>1$, i.e. $u>(\alpha')^{-1}(1)$. A disturbance that pushes $y$ upwards blows up immediately. A disturbance that pushes $y$ downwards sends $y$ to the stable branch. There, $y<\arg\max\beta$, so as long as $u-x\leq\max\beta$, i.e.\ $x\geq \arg\max\beta+u-(\alpha')^{-1}(1)$, equation \eqref{eq:sys2} gives $\tau\dot x=-x+y<-x+\arg\max\beta\leq-\big(u-(\alpha')^{-1}(1)\big)$, so $x$ decreases at a rate of at least $(u-(\alpha')^{-1}(1))/\tau$ until $u-x>\max\beta$, and the motif blows up by Theorem~\ref{thm:exact_internal}.
\end{proof}

Fig.~\ref{fig:thresholds} illustrates both the dynamic threshold of Theorem~\ref{thm:exact_internal} and the rheobase of Corollary~\ref{cor:rheobase}. It can be interpreted as follows. For a slow-varying input $u$, the state $x$ remains close to the equilibrium curve $x=\alpha(u)$, which stays below the dynamic threshold. Consequently, for a sufficiently slow input $u$, the motif does not explode until the input reaches the rheobase. In contrast, for a fast-varying input $u$, the state $x$ lags behind $\alpha(u)$, so a rapid increase in $u$ allows the dynamic threshold to be reached, leading to a blow-up of the motif below its rheobase. Whenever $u$ is a step signal, it changes instantaneously, before the internal state $x$ has time to respond, and it is straightforward to determine graphically whether the step will cross a threshold or not.

\section{Examples of the high-pass motif}\label{sec:examples}

\begin{figure}[t]
    \centering
    \includegraphics[width=0.85\linewidth]{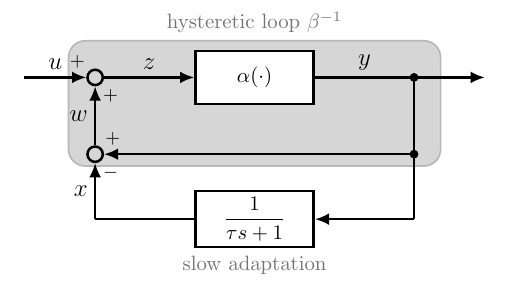}
    \caption{\small Collapsing the positive feedback loop of the high-pass motif yields the hysteresis-and-adaptation motif of the FitzHugh-Nagumo neuron model~\cite{SacrePHD}. After a causal regularization, the hysteretic loop $\beta^{-1}$ reads $\varepsilon \dot{y}=-\beta(y)+u-x$.}
    \label{fig:collapse}
\end{figure}

\subsection{FitzHugh-Nagumo neuron model} \label{sec:ex_FHN}

In this section, $\alpha$ is a sigmoid function with sufficiently high gain, such that its maximal slope exceeds one, and the motif \eqref{eq:motif} is excitable. The static positive feedback loop around $\alpha$, which we have interpreted so far as the direct feedthrough of the high-pass filter, can then be interpreted as an hysteretic loop, as it collapses with the sigmoid into a bistable element (Fig.~\ref{fig:collapse}). This bistable element may be interpreted as the inverse of the static path $\beta$ from the stable inverse motif introduced earlier (Fig.~\ref{fig:blocs}, Bottom). The remaining low-pass part of the high-pass filter then takes the role of a slow adaptation variable. Overall, the motif may also be interpreted as an hysteresis-and-adaptation interconnection, which is the structure of a FitzHugh-Nagumo neuron model~\cite{SacrePHD}.

\begin{figure}
    \centering
    \includegraphics[width=0.85\linewidth]{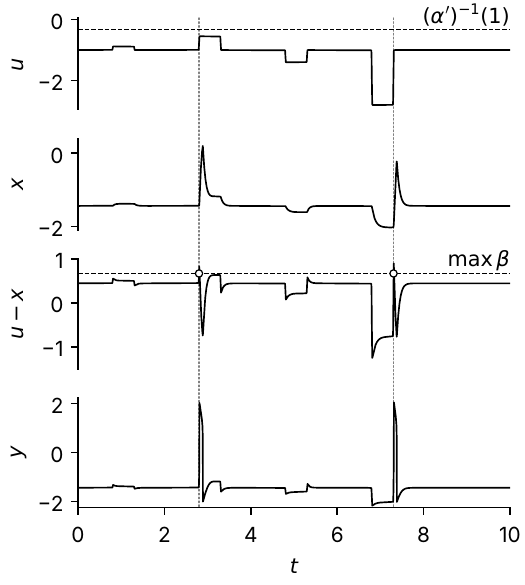}
    \caption{ \small Simulation of the FitzHugh-Nagumo neuron~\eqref{eq:FHN} with $\tau = 0.1$ and $\varepsilon = 0.001$. From top to bottom: time evolution of the input $u$, the internal state $x$, the thresholded quantity $u - x$, and the output $y$.}
    \label{fig:simu}
\end{figure}

Up to a positive time reparameterization of the fast dynamics, the causally regularized motif with its hysteretic loop collapsed reads as the following FitzHugh-Nagumo model:
\begin{subequations} \label{eq:FHN}
\begin{align}
    \varepsilon\dot y &= -\beta(y)-x+u, \\
    \tau\dot x &= -x+y.
\end{align}
\end{subequations}
Given the usual FitzHugh-Nagumo nonlinearity $\beta(y)=y^3/3-y$, our motif yields $\alpha^{-1}(y)=\beta(y)+y=y^3/3$, so $\alpha(z)=(3z)^{1/3}$. The two bifurcations of this FitzHugh-Nagumo model are located at $y=\pm1$. The rheobase is obtained by Corollary~\ref{cor:rheobase} at $u=(\alpha')^{-1}(1)=-1/3$. By Theorem~\ref{thm:exact_internal}, the spike is triggered from the lower branch exactly when $u-x>\max_{y\leq 0} \beta=\beta(-1)=2/3$. This threshold exactness is verified numerically in Fig.~\ref{fig:simu}.  The input $u$ stays below the rheobase at all times, yet the dynamic threshold is crossed twice: once by an excitatory pulse, once on release of an inhibitory pulse. This results in two output spikes.

\subsection{A stop-loss cascade model}

As mentioned in the introduction, the excitability property may not be specific to neurons, and it also may not always be desirable.

In the stock market, a stop-loss order refers to an automatic sell order of an asset triggered when the price of this asset falls below a predetermined level, typically defined as a percentage of a reference price. A stop-loss cascade is a phenomenon in which a sudden decline in the price of an asset triggers many stop-loss orders, initiating a self-reinforcing selling loop~\cite{Osler2005}. We show in this section that stop-loss cascades may be interpreted as an instance of our motif.

Since stop-loss orders are typically specified as percentages of a reference price and the price impact of sell orders is multiplicative, we work with logarithmic prices rather than prices. Let $p$ be the log-price of an asset and $\bar{p}$ its reference level. For simplicity, $\bar{p}$ is modeled as an exponentially weighted moving average of memory $\tau$.
\begin{equation} \label{eq:eco1}
    \tau\dot{\bar{p}}=-\bar{p}+p .
\end{equation}
In trading, the drawdown $d\triangleq \bar{p}-p$ of an asset stands for the loss of price of this asset with respect to a reference price $\bar{p}$. Given the dynamics of the reference price $\bar{p}$, we easily see that
\begin{equation} \label{eq:eco2}
    d=-H(s)p, 
\end{equation}
so the drawdown of an asset can be roughly interpreted as its high-pass-filtered log-price. A slow decline in $p$ is tracked by the moving average reference $\bar{p}$ and produces no drawdown, whereas a fast decline in $p$ will be registered by $d$.

The drawdown signal $d$ is tracked by stop-loss orders, which liquidate a position once its loss exceeds a preset tolerance. We denote by $\Phi(d)$ the volume of selling orders that are triggered for a drawdown signal of value $d$ or below. Since $\Phi$ is a cumulative distribution of tolerances across all stop-loss orders in the market, it is increasing. A selling order depresses the log-price $p$ of an asset by $\lambda\Phi(d)$ below an exogenous log-price $p_0$, where $p_0$ represents the log-price the asset would follow in the absence of forced selling:
\begin{equation}  \label{eq:eco3}
    p_0-p=\lambda\Phi(d).
\end{equation}
Combining \eqref{eq:eco1}, \eqref{eq:eco2} and \eqref{eq:eco3}, we exactly recover the high-pass motif of excitability~\eqref{eq:motif}. The output $y\triangleq p_0-p = \lambda\Phi(d)$ is the log-price decline caused by the cascade, the input $u\triangleq-H(s)p_0$ is the part of the exogenous decline that is too abrupt to be absorbed by the moving-average reference, the feedback $w\triangleq H(s)(p_0-p)$ is the observed drawdown minus the part attributable to the exogenous decline, the variable $z\triangleq d$ is simply the drawdown signal, and the activation function is given by $\alpha\triangleq\lambda\Phi$. The regularization time constant $\varepsilon$ corresponds to the execution time of sell orders, which is typically negligible compared to the moving-average time constant $\tau$.

The system becomes \textit{excitably fragile} whenever $\alpha'(z)=\lambda\Phi'(z)>1$, which depends on both the distribution of volume of selling orders $\Phi$, and the multiplicative price impact of a sell order $\lambda$. Under conditions of excitable fragility, a stop-loss cascade may be triggered by a fast shock, for example by the rapid spread of false information. The sudden price collapse is then driven primarily by the excitability property rather than by any fundamental change in the value of the asset~\cite{kirilenko2017flash}.

\subsection{A social contagion model}

The same excitability structure could be governing the spread of a behavior in a population, for example: a rumor, a protest, a fashion trend, or panic buying. Let $p$ denote the fraction of the population currently engaged in a behaviour, and let $\bar p$ denote the fraction of the population that people are accustomed to seeing participate in it. We model $\bar p$ as a moving average of memory $\tau$, so that \eqref{eq:eco1} holds. Because of social habituation, individuals may respond more strongly to a recent increase in participation in a behavior than to its absolute level. The relevant signal for social contagion could therefore be the perceived salience of a behaviour, which is, once again, captured by a high-pass filtering mechanism: $w\triangleq p-\bar p=H(s)p$. The perceived salience $w$ is large under a sudden increase of $p$, but vanishes under steady participation.

Following~\cite{Granovetter1978}, we consider that each individual starts to participate in the behavior once the perceived salience exceeds a personal tolerance, so that $p=\Phi(z)$, where $\Phi$ is the cumulative distribution of tolerances and $z=u+w$ is the total salience, $u$ being its exogenous part, e.g. a surge in media coverage. Setting $y\triangleq p$ and $\alpha\triangleq\Phi$ recovers once again the high-pass motif~\eqref{eq:motif}. Here, the regularization time constant $\varepsilon$ corresponds to the time taken to join the behavior, which we again consider negligible against $\tau$. The excitability condition $\alpha'(z)=\Phi'(z)>1$ states that one additional unit of perceived salience recruits enough participants to generate more than one unit in return.

\section{Conclusions and perspective}\label{sec:conclusion}

We introduced a minimal motif for excitability based on a high-frequency positive feedback loop. Using the stable inverse of this motif, we derived an explicit dynamical threshold separating signals that trigger excitation and cause the motif to blow up from those that maintain stability. This threshold is a simple inequality on the difference between the instantaneous input and the motif's internal memory. We illustrated the relevance of this framework with three examples: the FitzHugh-Nagumo model of neuronal dynamics, where the threshold provides a direct characterization of excitability, and a stop-loss cascade model in the stock market, as well as a social contagion model, showing that similar excitability can arise in rate-induced tipping systems.

On the conceptual side, the current work remains to be generalized to a setting in which the high-frequency positive feedback is implemented by a linear, $n$th-order, high-/band-pass filter. Under this setting, a key objective would be to develop a unified framework for excitability that connects its classical geometric interpretation in terms of bifurcations~\cite{Izhikevich2006}, its recently introduced energy-based interpretation~\cite{Sepulchre2025}, and a yet-to-be-established frequency-domain perspective.

On the practical side, our proposed threshold remains to be applied to real-world systems spanning neurobiology and neuromorphic engineering, as well as the critical systems studied in the tipping-point literature. For the latter specifically, our motif remains to be identified within existing models to establish its relevance beyond the stop-loss cascade model and the social contagion model considered in this paper.

\bibliographystyle{ieeetr}
\bibliography{refs}

\end{document}